# My friend Alex Müller

Erio Tosatti[1,2,3,*]

[1]International School for Advanced Studies (SISSA), Via Bonomea 265, 34136 Trieste, Italy

[2]International Centre for Theoretical Physics (ICTP), Strada Costiera 11,34151 Trieste, Italy

[3]Consiglio Nazionale delle Ricerche - Istituto Officina dei Materiali (CNR-IOM), c/o SISSA, Trieste, Italy

(Dated: August 21, 2023)

## Abstract

Alex, the main discoverer of high $T_c$ superconductivity, was also a dear friend. Here I offer a few frank anecdotes, possibly inaccurate in some details but heartfelt and accurate in the substance, as a personal tribute to our friendship.

I first met Alex at a meeting in the Swiss Alps in the 1970s. The meeting subject was mostly on spectroscopy of crystalline solids, a hot topic at the time. My dear friend and collaborator Gunther Harbeke of RCA Zurich Laboratories had brought me there. The community was European but rather Zurich-centered. Alex, head scientist of IBM Zurich Laboratories in Rueschlikon, was there too, with a lot of physics on his hands. With Gunther we presented results on layered materials. Incidentally, Gunther and Dave Greenaway had shown me how they deftly peeled layers off jewel quality Transvaal graphite (layers nowadays called graphene) with Scotch tape – in 1968! Alex instead had recently shown, among other things, how his $Fe(3+)$ EPR spectra could disclose the universality of a 3D displacive structural transition – specifically I believe it was the antiferrodistortive transition of $SrTiO_3$ at 105 K. My own expertise at the time, optical properties of semiconductors and insulators, was not at all in the same field. I was young, and had only to learn from him, and from everybody there. Despite that scientific gap, Alex and I resonated on the human side. Having spent time during his youth, he told me, in Cantone Grigioni, Alex spoke decent Italian, and enjoyed practising it. He already had, and later continued, a steady and lively interaction with the very distinguished colleague Attilio Rigamonti of University of Pavia, with whom he even organized an influential Varenna School [1]. During the meeting we talked about many things outside of physics, and of course inside physics too. From the excitons in layered materials I was working on at the time, to his beloved ferroelectrics and strontium titanate. A subject I knew nothing about, but which already constituted the main scientific thread of his life. I was duly impressed, both on the human and scientific side. But after that isolated occasion our lives continued their course along different routes.

Ten years later, now established in Trieste, I was ready for a sabbatical, and Zurich seemed the ideal place. Full of good physics, especially at IBM where STM had just been invented by Gerd Binnig and Heini Rohrer, and where Toni Schneider had generously invited me to join the theory section. Also in Zurich there were the RCA Laboratories with Gunther Harbeke, with whom collaboration and friendship were still lively and fruitful. Zurich was top choice family-wise too, for I was planning to bring my wife Rita with little Valentino, three, and Susanna, two. Our prospective housing in the Irchel neighborhood of the upper Scheucherstrasse, then populated by down-to-earth families who raised as many kids as their apartments could contain, looked perfect for family life. We might never have really pursued our planned third child Giulio, had we not spent that year in the Scheucherstrasse. The nearby Irchel hill, a bonus I was to discover later, is also the best place imaginable for kite flight – an art which I practiced pretending to do it for our kids.

So I went to Zurich in October 1984. And after three months with Gunther at RCA, I found myself from January 1985 with Alex at IBM Rueschlikon. In the same corridor, essentially next door. Thus our communality restarted. We met each other's families, in particular we met Inge and enjoyed visiting their house. As a practical man, Alex abundantly helped me out in menial, but urgent matters. For example, that of 1984-85 was a classic Swiss winter. After the standard Hochnebel in late fall, Zurich was full of snow and ice everywhere. That was a lot of fun; but in the mornings my asthmatic Fiat could not start, the heavy diesel engine dead frozen. Alex, gleeful at my dismay, got me readily out of trouble with his stick-shaped electric boiler. In the lab we broadly discussed physics around disparate subjects, in particular graphite intercalation compounds and more. We generally would go for lunch together, every time with a lot of freewheel chatting.

On one of these lunches, in the sunny outdoors IBM cafeteria, I recall boasting with him the magic of quantum melting (QM) of crystal lattices. The fantastic work of Laughlin on the fractional quantum Hall effect (FQHE), due to QM of a 2D Wigner lattice once confined in the lowest Landau level, was very recent, and had impressed many of us. That was yet another example of that magic, after liquid He resulting from QM of fcc He, superconductivity from QM of charge-density-waves, and Anderson's 1973 Resonating Valence Bond (RVB, nowadays called a spin liquid) from QM of frustrated triangular spin 1/2 antiferromagnet. Every case of QM so far had, I stressed, some remarkable – often super – property! So I wondered, what about QM of a ferroelectric or antiferroelectric, with moments losing their static long ranged orientation owing to quantum fluctuations ? Could such a rotational QM exist for electric dipole moments? And if so, might it have any unexpected property? Alex jumped in his chair " Erio" he said, "I already have an example of such a state – it's incipient ferrolelectric strontium titanate!" In brief, $SrTiO_3$ tries to go ferroelectric like $BaTiO_3$ upon cooling. But because Sr is a smaller ion than Ba, the $TiO_6$ octahedral cage is tighter, and the classical extrapolated Curie temperature is an order of magnitude smaller, down from hundreds to tens of K. Once these low temperatures are reached in $SrTiO_3$, quantum mechanics is no longer negligible. The dielectric constant grows up to an unheard-of 20,000 upon cooling, but then levels off rather than diverging critically at a classical ferroelectric onset. Ferroelectricity is killed precisely by quantum fluctuations below 37 K, the temperature where Alex and H. Burkard placed the onset of quantum paraelectricity [2]. The emerging curiosity out of that lunchtime chat was thus whether or not some new, yet unidentified quantum order parameter was to accompany QM. In which case its onset might show up as a phase transition anomaly upon cooling across 37 K. Guess how that anomaly could be looked for? Of course by $Fe(3+)$ EPR!

There were of course also good reasons for doubt. Examples of QM systems without broken symmetries or arcane order parameters do exist. They occur in systems possessing, or developing, a spectral excitation gap, implying an uneventful thermal evolution upon cooling or heating. One example was the famous Ising model in transverse field, solved among others by De Gennes in 1963 [3]; another was the FQHE itself, solved by Laughlin in 1983 [4]. In addition, after throwing the stone, I had no precise idea of what new order parameter, if any, might appear in a quantum paraelectric. Impermeable to all doubts, Alex was unstoppable at that point – he just wanted to see if experiment would or not reveal a phase transition in $SrTiO_3$ near 37 K. He ran to Walter Berlinger, his faithful inseparable technician, and saddled him with the task.

Soon came the answer – and bingo! there was indeed a weak but unmistakable dip singularity at 37 K in both a and b parameters of the EPR spin hamiltonian fit to Walter's data. The dip suggested a lattice expansion: but Alex knew for sure that $SrTiO_3$ has no such static expansion at 37 K. The dip singularity could even be followed upon changing orientation and pressure. So here was, apparently crystal clear, the evidence of our coveted phase transition! Alex analysed the data and promptly wrote his part of the paper, asking me to add the theory. Alas, what theory to proclaim after setting all this machinery into motion I did not quite know. To increase my embarrassment, came in the meanwhile Alex's Nobel prize for the discovery of HiTc superconductivity. He and Georg Bednorz had been quietly chewing on their masterpiece work on cuprate perovskites precisely while I was next door to them in Rueschlikon. They had kept it top secret, probably pending the approval of skeptical IBM headquarters. Now, having proven himself right against all odds on cuprate perovskites, Alex was once again ready with a novel quantum state, a result which he privately named "another high flier".

Sweetest to Alex was the prospective discovery of an exciting novel state precisely in his beloved $SrTiO_3$. An amusing anecdote Alex told me which I cannot resist recounting describes a small event that took place in Stockholm. During the Nobel ceremonies, his wife Inge donned such an incredible diamond parure to stir considerable curiosity. Alex's absolute delight was to explain that the crystal in Inge's jewels was not diamond, it was strontium titanate!

With no theory to offer I managed, partly helped by the HiTc hype surrounding Alex, to resist a few years his urge to publish the novel quantum state of $SrTiO_3$--meanwhile fruitlessly racking my brains for an explanation. Eventually, the experimental result was undeniable, and Alex's insistence to publish fully justified. We thus penciled as a possible explanation the handwaving idea of an incommensurate modulated state driven by acoustical phonon softening at finite k-vector. The superlattice could then quantum melt due to excessive quantum fluctuations, similarly to the helium lattice. Some nondescript off-diagonal long range order could then materialize, with the superlattice acoustical phonons turning roton-like, and gapped. In this form the paper was out, once again on Zeitschrift fur Physik [5], sort of mimicking that of Alex and Georg on cuprates.[6] The work stirred interest, skepticism, and some follow-ups, of course nothing like the cuprates. Later, with my PhD student Roman Martonak, we published a thorough Quantum Monte Carlo study of a simplified but we thought pertinent Hamiltonian [7] which confirmed the non-existence of off-diagonal long range order, and the presence of an excitation gap. At least in that approximation, everything was not fundamentally different from Ising in transverse field, or from FQHE – alas offering no explanation for the EPR dips. Alex of course did not like our conclusions, and our views on the significance of the EPR results parted.

The right qualitative explanation came I believe several years later by the late Jim Scott and collaborators.[8] What happens in SrTiO3 at 37 K is that the ferroelectric TO mode, an optically active phonon which softens upon cooling, crosses the antferrodistortive mode, an optically inactive phonon which hardens upon cooling, as temperature is well below the famous 105 K antiferrodistortive phase transition. The two modes have different symmetry, and would in a harmonic world cross uneventfully. However, anharmonicity can play here an unexpected trick. A third mode, acoustical and of very small frequency and wavevector, can be created when the modes are about to cross. The symmetry of that acoustical mode being low (because its k-vector differs from zero) it can couple the two orthogonal optical modes when their frequencies are close but not exactly coincident; and that coincidence happens, Scott et al. noted, precisely near 37 K. At that temperature the crystal can thus be pervaded by a new superlattice periodicity – a bit like in our handwaving story, but now consisting of a time and temperature dependent superlattice-like modulation. A longitudinal acoustical mode has, at short range, similar features to a lattice expansion/contraction: and that may be what gives rise the EPR a and b parameters dip. That makes on the whole an interesting even if not so glorious piece of physics, incidentally one that so far as I know still waits to be demonstrated theoretically. If that is the mechanism, the EPR dips do not actually signal the onset of the quantum order parameter of a novel state associated with the QM of ferroelectricity. More recent work actually explores subtler features of the quantum paraelectric state of SrTiO3 at much lower temperatures, also associated with electrons from oxygen vacancies as described by Peter Littlewood and others [9] but that is a different story. And maybe there is indeed some magic in paraelectric SrTiO3 after all; but that is most likely not signaled by the 37 K EPR anomaly.

After 1986 Alex and I did of course a lot of talking about HiTc superconductors too. His results on cuprate perovskites boldly challenged all six rules proclaimed prior to 1986 by Bernd Matthias, another Swiss and famous Bell Labs experimentalist, great guru of super-conductivity : high symmetry is good; cubic symmetry is best; high density of electronic states is good; stay away from oxygen; stay away from magnetism; stay away from insulators; stay away from theorists. Alex was very fond of the Jahn Teller effect, which is indeed a source of molecular singlet electron pairing. Following that hunch- also supported by strong theorists friends of Alex like Harry Thomas - Gerd Binnig, then the freshest entry in IBM Rueschlikon, had, with his collaborators and stimulated by Alex, then head of the Laboratory, discovered and qualified superconductivity in marginally Nb-doped $SrTiO_3$ [10]. Consistent with that, Alex continued to believe in a Jahn Teller, phononic or bipolaronic origin of superconductivity in the cuprates. Resisting Anderson's idea, which time seems now to vindicate more and more, that these materials are strongly correlated, basically just doped RVB Mott insulating states.

When, in the summer of 1987, Stig Lundquist (then also secretary of Physics Nobel Committee), myself, Mario Tosi, and Yu Lu organized in ICTP an Adriatico Conference on High Temperature Superconductivity [11] I was also in charge of convincing Alex, tired after the US "Woodstock of Physics" of previous March, to come and speak up. I explained to him that Trieste was at least as important as New York's Woodstock venue when it came to inform the world, including Europe and beyond, of his mind-boggling results, contrary to at least five Bernd's "rules". In the end I did convince him – most likely not by my arguments, but rather because coming to Trieste would give him the thrill to drive his Jaguar across the Alps!

The last time I saw Alex and Inge was a few years ago. In good shape, living in a super-comfortable residence in the Zurich outskirts, a place with theater, restaurants, and facilities for both leisure and quiet thinking. I, a Modenese, made them the last present of my own Traditional Balsamic Vinegar. In turn, I was treated with much smiling, plenty of brotherly private confidences, as well as with a splendid meal. Characteristically, Alex said he had stopped thinking about physics – except that by talking to him I could see that in reality that was not true…

[*] tosatti@sissa.